\documentclass[aps,prd,groupedaddress,showpacs,nofootinbib]{revtex4}

\usepackage{graphicx,dcolumn,bm,amssymb,amsmath,latexsym,amsfonts,footnote}

\begin{document}

\newcommand{\nonu}{\nonumber}
\newcommand{\sm}{\small}
\newcommand{\noi}{\noindent}
\newcommand{\npg}{\newpage}
\newcommand{\nl}{\newline}
\newcommand{\bp}{\begin{picture}}
\newcommand{\ep}{\end{picture}}
\newcommand{\bc}{\begin{center}}
\newcommand{\ec}{\end{center}}
\newcommand{\be}{\begin{equation}}
\newcommand{\ee}{\end{equation}}
\newcommand{\beal}{\begin{align}}
\newcommand{\eeal}{\end{align}}
\newcommand{\bea}{\begin{eqnarray}}
\newcommand{\eea}{\end{eqnarray}}
\newcommand{\bnabla}{\mbox{\boldmath $\nabla$}}
\newcommand{\univec}{\textbf{a}}
\newcommand{\VectorA}{\textbf{A}}
\newcommand{\Pint}

\title{Corotating binary systems of identical Kerr-Newman black holes}

\author{I. Cabrera-Munguia$^{1,}$\footnote{icabreramunguia@gmail.com}, Etevaldo dos Santos Costa Filho$^{2}$, H\'ector H. Hern\'andez$^{3}$, and David V\'azquez-Valdez$^{1}$ }
\affiliation{$^{1}$Departamento de F\'isica y Matem\'aticas, Universidad Aut\'onoma de Ciudad Ju\'arez, 32310 Ciudad Ju\'arez, Chihuahua, M\'exico\\
$^{2}$Instituto de F\'isica de S\~{a}o Carlos-USP, 13566-970 S\~{a}o Carlos, S\~{a}o Paulo, Brazil\\
$^{3}$Universidad Aut\'onoma de Chihuahua, Facultad de Ingenier\'ia, Nuevo Campus Universitario, Chihuahua 31125, M\'exico}


\begin{abstract}
In the present paper binary configurations of identical corotating Kerr-Newman black holes separated by a massless strut are derived and studied. After solving the axis conditions and establishing the absence of magnetic charges in the solution, one gets two 4-parametric corotating binary black hole models endowed with electric charge, where each source contains equal/opposite electric charge in the first/second configuration. Since the black hole horizons are given by concise expressions in terms of physical parameters, all their thermodynamical properties satisfying the Smarr relation for the mass are also obtained. We discuss the physical limits of both models.
\end{abstract}

\maketitle

\vspace{-0.5cm}
\section{Introduction}
\vspace{-0.3cm}
Recent results on gravitational waves detection \cite{LIGO} open new expectations on the search of interacting binary black hole (BH) models that might be helpful to analyze and study this physical phenomenon in an exact form, since until this day numerical relativity has been the main tool to treat the process of binary BH mergers. Regarding the last point, in stationary spacetimes, simplified models of binary systems have been taken into account since the early days of general relativity, perhaps the most famous is that one described by the Bach-Weyl solution \cite{BachW} which illustrates two arbitrary Schwarzschild BHs interacting due to their gravitational attraction. In Einstein vacuum systems, the double-Kerr solution \cite{KramerNeugebauer} of Kramer and Neugebauer is very helpful to describe unequal binary configurations of interacting BHs, where the nonlinear superposition of the fields of each Kerr BH is carried out after solving analytically the axis conditions, permitting that both sources are held apart by a conical singularity \cite{BachW,Israel}. The solving of the axis conditions had been one of the main highly complicated problems to study dynamical and thermodynamical aspects of two interacting Kerr BHs, which fortunately has been concluded recently in \cite{Cabrera2018}. Naturally, one may have in mind the possibility of extending this result by adding the electromagnetic field. However, the bad thing is that such a process increases enormously the complexity of finding exact results, since the axis conditions must be solved in combination with the condition that avoid the presence of magnetic charges in the solution, with the aim to determine binary configurations of Kerr-Newman BHs separated by a massless strut (conical singularity). The last point suggests us the idea of treating cases of identical BH configurations, due to their more symmetric character.

This letter pursues the main objective of extending the earlier results provided in \cite{Costa,CCLP} in relation to identical corotating BHs, where now the sources containing aligned spins will be endowed with electric charges. In this work, we derive two $4$-parametric binary models of corotating Kerr-Newman BHs where the first of them contains equal electric charges, while the second one carries opposite electric charges. In addition, all the physical limits and thermodynamical properties of both models are well defined by concise expressions in terms of arbitrary Komar parameters \cite{Komar}. It is also included a concise metric in the extreme limit case of BHs, where are obtained simple expressions for the force related to the strut and area of the horizon during the touching limit, extending the recent result of \cite{Ciafre-Rodriguez}.

\vspace{-0.5cm}
\section{ The asymptotically flat exact solution}
\vspace{-0.3cm}
Within the context of exact solutions it is well-known that stationary axisymmetric spacetimes can be described by means of the line element
\cite{Papapetrou}
\vspace{-0.1cm}
\be ds^{2}=f^{-1}\left[e^{2\gamma}(d\rho^{2}+dz^{2})+\rho^{2}d\varphi^{2}\right]- f(dt-\omega d\varphi)^{2}.
\label{Papapetrou}\ee

\vspace{-0.1cm}
\noi where the metric coefficients $f,\, \omega$ and $\gamma$ depend only on cylindrical coordinates $(\rho,z)$. In this sense, Ernst formalism reduces the Einstein-Maxwell field equations into a new coupled system \cite{Ernst}
\vspace{-0.1cm}
\bea \begin{split}  \left({\rm{Re}} {\cal{E}}+|\Phi|^{2}\right)\Delta{\cal{E}}&=(\bnabla{\cal{E}}+
2\bar{\Phi}\bnabla \Phi)\cdot \bnabla {\cal{E}}, \\
 \left({\rm{Re}}{\cal{E}}+|\Phi|^{2}\right)\Delta \Phi&=(\bnabla{\cal{E}}+
2\bar{\Phi}\bnabla\Phi)\cdot\bnabla\Phi, \label{Ernst} \end{split} \eea

\vspace{-0.1cm}
\noi where ${\cal{E}}=f-|\Phi|^{2}+i\Psi$ and $\Phi=-A_{4}+i A'_{3}$ are the Ernst complex potentials. Any explicit knowledge of $({\cal{E}},\Phi)$ provides the above stationary metric Eq.\ (\ref{Papapetrou}) after solving a complicated set of differential equations:
\vspace{-0.1cm}
\begin{align}
4\gamma_{,\rho}&=\rho f^{-2} \left[|{\cal{E}}_{,\rho}+
2\bar{\Phi}\Phi_{,\rho}|^{2} -|{\cal{E}}_{,z}+ 2\bar{\Phi}\Phi_{,z}|^{2}\right] - 4\rho f^{-1}(|\Phi_{,\rho}|^{2}- |\Phi_{,z}|^{2}),\nonu\\
2\gamma_{,z}&=\rho f^{-2}{\rm{Re}} \left[({\cal{E}}_{,\rho}+
2\bar{\Phi}\Phi_{,\rho})(\bar{{\cal{E}}}_{,z}+ 2\bar{\Phi}\Phi_{,z})\right]-4\rho f^{-1} {\rm{Re}(\bar{\Phi}_{,\rho}\Phi_{,z})},\nonu\\
\omega_{,\rho}&=-\rho f^{-2}{\rm{Im}}( {\cal{E}}_{,z}+ 2 \Phi\bar{\Phi}_{,z}),\qquad
\omega_{,z}=\rho f^{-2}{\rm{Im}}( {\cal{E}}_{,\rho}+ 2 \Phi\bar{\Phi}_{,\rho}).
\label{metricfunctions}\end{align}

\vspace{-0.1cm}
\noi The difficult task to obtain exact solutions from Ernst's equations can be achieved by using Sibgatullin's method (SM) \cite{Sibgatullin}; a modern generation technique of exact solutions in stationary axisymmetric Einstein-Maxwell spacetimes, which is based on the soliton theory. According to SM, we must begin with the next representation of the Ernst potentials on the upper part of the symmetry axis (the axis data) \cite{RMJ}:
\vspace{-0.1cm}
\be e(z)=1+\sum_{i=1}^{2}\frac{e_{i}}{z-\beta_{i}}, \qquad f(z)=\sum_{i=1}^{2}\frac{f_{i}}{z-\beta_{i}},\label{generalernst}\ee

\vspace{-0.1cm}
\noi being $e(z):={\cal{E}}(\rho=0,z)$ and $f(z):={\Phi}(\rho=0,z)$. In addition, the above axis data is depicted by six arbitrary complex constants contained inside of the set $\{e_{i},f_{i}, \beta_{i}\}$, $i=1,2$. Since the SM provides the Ernst potentials, and therefore, the full metric in the whole spacetime once we adopt a specific form of the axis data, it is necessary to give it first a more physical representation in order to gain more insight and simplicity at the moment of studying physical and dynamical aspects of binary systems. To accomplish such a task we begin with the characteristic equation
\vspace{-0.1cm}
\be e(z) + \bar{e}(z)+ 2 f(z) \bar{f}(z)=0,\label{characteristic}\ee

\vspace{-0.1cm}
\noi whose roots $\alpha_{n}$, for $n=\overline{1,4}$, define the location of the sources on the symmetry axis. By placing the axis data Eq.\ (\ref{generalernst}) into this characteristic equation we have
\vspace{-0.1cm}
\begin{align} &2+ \sum_{i=1}^{2}\left(\frac{e_{i}}{z-\beta_{i}}+ \frac{\bar{e}_{i}}{z-\bar{\beta}_{i}}\right)
+2 \sum_{i,j=1}^{2} \frac{f_{i}\bar{f}_{j}}{(z-\beta_{i})(z-\bar{\beta_{j}})}
=\frac{2\prod_{n=1}^{4}(z-\alpha_{n})}{\prod_{i=1}^{2}(z-\beta_{i})
(z-\bar{\beta}_{i})},
\label{characteristicII} \end{align}

\vspace{-0.1cm}
\noi where after performing a partial fraction decomposition it is possible to obtain the relations
\vspace{-0.1cm}
\begin{align} e_{1}&=\frac{2 \prod_{n=1}^{4}(\beta_{1}-\alpha_{n})}
{(\beta_{1}-\beta_{2})(\beta_{1}-\bar{\beta}_{1})(\beta_{1}-\bar{\beta}_{2})}-\sum_{i=1}^{2} \frac{2f_{1}\bar{f}_{i}}{\beta_{1}-\bar{\beta}_{i}},\qquad
e_{2}=\frac{2 \prod_{n=1}^{4}(\beta_{2}-\alpha_{n})}
{(\beta_{2}-\beta_{1})(\beta_{2}-\bar{\beta}_{1})(\beta_{2}-\bar{\beta}_{2})}-\sum_{i=1}^{2} \frac{2f_{2}\bar{f}_{i}}{\beta_{2}-\bar{\beta}_{i}}, \label{thees}
\end{align}

\vspace{-0.1cm}
\noi which allows us to change the set of parameters $\{e_{i},f_{i}, \beta_{i}\}$ by the new ones $\{\alpha_{n},f_{i}, \beta_{i}\}$. It follows that the first Simon's multipole moments \cite{Simon} like the total mass of the system $M$, NUT charge $J_{0}$ \cite{NUT}, as well as the total electromagnetic charge $Q+iB$ can be computed from the above axis data Eq.\ (\ref{generalernst}) via the Hoenselaers-Perj\'es procedure \cite{HP,Sotiriou}, having
\vspace{-0.2cm}
\be -\frac{e_{1}+e_{2}}{2}=M + i J_{0}, \qquad f_{1}+f_{2}=Q+iB.\label{total}\ee

\vspace{-0.1cm}
Additionally, the total angular momentum $J$, and electric/magnetic dipole moment  $\mathcal{Q}_{o}/\mathcal{B}_{o}$ are given by
\vspace{-0.1cm}
\begin{align} &{\rm {Im}}\left[\left(\frac{e_{1}+e_{2}}{2}\right)^{2}-\frac{ e_{1}\beta_{1}+e_{2}\beta_{2}}{2}\right]=J, \qquad -\frac{(e_{1}+e_{2})(f_{1}+f_{2})}{2}+f_{1}\beta_{1}+f_{2}\beta_{2}=\mathcal{Q}_{o}+i\mathcal{B}_{o},
\label{total2}\end{align}

\vspace{-0.1cm}
\noi and after placing Eq.\ (\ref{total}) into the second formula of Eq.\ (\ref{total2}) it might be possible to get the following expressions:
\vspace{-0.1cm}
\be f_{1,2}=\pm \frac{-(Q+iB)(M+iJ_{0}+\beta_{2,1})+\mathcal{Q}_{o}+i\mathcal{B}_{o}}{\beta_{1}-\beta_{2}}. \label{theefs}\ee

\vspace{-0.1cm}
On the other hand, the substitution of Eq.\ (\ref{thees}) inside of the left-hand side of Eq.\ (\ref{total}) leads to the relation for the total mass
\vspace{-0.2cm}
\be \beta_{1}+ \beta_{2}+ \bar{\beta}_{1}+\bar{\beta}_{2}-\sum_{n=1}^{4}\alpha_{n}=-2M, \label{themass}\ee

\vspace{-0.1cm}
\noi whereas the choice of the suitable parametrization
\vspace{-0.1cm}
\be \alpha_{1,2}=\frac{R}{2}\pm \sigma_{1}, \qquad \alpha_{3,4}=-\frac{R}{2}\pm \sigma_{2},
\label{thealphas}\ee

\vspace{-0.1cm}
\noi reduces one parameter in the general solution, since $\sum_{n=1}^{4}\alpha_{n}=0$. In this case $R$ plays the role of the relative distance among the sources, where $\sigma_{i}^{2}\geq 0$ defines BHs while $\sigma_{i}^{2}<0$ is referring to hyperextreme sources. By adopting the simple redefinitions $\mathcal{Q}_{o}=q_{o}-B(\mathfrak{q}+J_{0})$, $\mathcal{B}_{o}=b_{o}+Q(\mathfrak{q}+J_{0})$, and $\beta_{1}+\beta_{2}=-M+i\mathfrak{q}$, the above-mentioned Eq.\ (\ref{theefs}) turns out to be
\vspace{-0.1cm}
\be f_{1,2}=\pm \frac{(Q+iB)\beta_{1,2}+q_{o}+i b_{o}}{\beta_{1}-\beta_{2}}.\ee

\vspace{-0.1cm}
Therefore, the problem of representing the axis data with a more physical appearance is accomplished with the choice of $\beta_{i}$ that satisfies Eq.\ (\ref{themass}) as follows:
\vspace{-0.1cm}
\be \beta_{1,2}=\frac{-M+i\mathfrak{q} \pm \sqrt{p+i \delta}}{2},\ee

\vspace{-0.1cm}
\noi where a few more trivial redefinitions given by
\vspace{-0.2cm}
\begin{align} p&=R^{2}+M^{2}-\mathfrak{q}^{2}-2\Delta_{o} +2\left(\epsilon_{1}-\frac{\epsilon_{2}R-\mathfrak{q}{\rm S}_{1}+2(q_{o}Q+b_{o}B)}{M}\right),\nonu\\
\delta&=-2(2P_{2}+M\mathfrak{q}),\quad {\rm S}_{1}=P_{1}+P_{2},\quad
\epsilon_{1,2}= \sigma_{1}^{2} \pm \sigma_{2}^{2}, \quad
\Delta_{o}= M^{2}-Q^{2}-B^{2}-\mathfrak{q}^{2},
\label{thepdelta}\end{align}

\vspace{-0.1cm}
\noi permit us to demonstrate that the Ernst potentials given by Eq.\ (\ref{generalernst}) acquire the final aspect
\vspace{-0.2cm}
\begin{align}
{\cal E}(0,z)&=\frac{z^{2}-[M + i(\mathfrak{q}+2J_{0})]z +\mathcal{P}_{+}+i P_{1} -2iJ_{0}  \Big[M-i\mathfrak{q}+\frac{P_{2}}{\mathfrak{q}}\Big]}{z^{2} + (M -i\mathfrak{q})z + \mathcal{P}_{-} + i P_{2}}, \qquad \Phi(0,z)=\frac{(Q+iB)z+\mathfrak{q}_{o}}{z^{2} + (M -i\mathfrak{q})z + \mathcal{P}_{-} + i P_{2}}, \nonu \\
\mathcal{P}_{\pm}&= \frac{M(2\Delta_{o}-R^{2})-2\left[M\epsilon_{1}\pm \big(\epsilon_{2}R-\mathfrak{q}{\rm S}_{1}+2(Q q_{o}+ B b_{o})\big)\right]}{4M}, \quad \mathfrak{q}_{o}=q_{o}+ib_{o},
\label{ernstaxiselectrogeneral}\end{align}

\vspace{-0.1cm}
\noi while the NUT charge and total angular momentum are simplified as
 \vspace{-0.2cm}
\begin{align} J_{0}&=\frac{\mathfrak{q}}{8M^{2}} \left( \frac{N}{\mathfrak{q}^{2}\mathcal{P}_{-}+P_{2}(P_{2}+M\mathfrak{q})} \right), \qquad
J=M \mathfrak{q}-\frac{P_{1}-P_{2}}{2}+J_{0}\left(2M+\frac{P_{2}}{\mathfrak{q}}\right),  \nonu \\
N&=M^{2}\left\{4(P_{1}P_{2}+|\mathfrak{q}_{o}|^{2})+(R^{2}-\Delta_{o})
(2\epsilon_{1}-\Delta_{o})+\epsilon_{2}^{2}\right\} -\left[\mathfrak{q}{\rm S}_{1}-\epsilon_{2}R-2(Q q_{o}+ B b_{o})\right]^{2}, \quad |\mathfrak{q}_{o}|^{2}=q_{o}^{2}+b_{o}^{2}.
\label{NUTcharge}\end{align}

\vspace{-0.1cm}
\noi It is worthwhile to stress the fact that in the absence of the electromagnetic field Eq.\ (\ref{ernstaxiselectrogeneral}) reduces to the axis data derived in Ref.\ \cite{Cabrera2018} for vacuum systems, which have been very fit to treat unequal configurations of interacting BHs. With the main purpose to treat binary configurations composed by identical sources, we just make $\sigma_{1}=\sigma_{2}=\sigma$, $P_{1}=-P_{2}=\delta$ as well as the following changes $\{M,\mathfrak{q},Q,B,q_{o},b_{o}\} \rightarrow  \{2M,2\mathfrak{q},2Q,2B,2q_{o},2b_{o}\}$, to obtain
\vspace{-0.1cm}
\begin{align}
{\cal E}(0,z)&=\frac{z^{2}-2[M + i(\mathfrak{q}+2J_{0})]z +P_{+}+i\delta -8iJ_{0} \Big[M-i\mathfrak{q}-\frac{\delta}{4\mathfrak{q}}\Big]}{z^{2} + 2(M -i\mathfrak{q})z + P_{-}-i \delta}, \qquad \Phi(0,z)=\frac{2(Q+iB)z+2\mathfrak{q}_{o}}{z^{2} + 2(M -i\mathfrak{q})z + P_{-}-i \delta}, \nonu \\
P_{\pm}&= 2\Delta_{o}-R^{2}/4-\sigma^{2} \mp 2(Q q_{o}+ B b_{o})/M,
\label{ernstaxiselectro}\end{align}

\noi where now $2M$ and $2(Q+iB)$ represent the total mass and total electromagnetic charge of the system, respectively. In addition, the electric/magnetic dipole moment $\mathcal{Q}_{o}/\mathcal{B}_{o}$ is given by
\vspace{-0.1cm}
\be \mathcal{Q}_{o}=2q_{o}-4B(\mathfrak{q}+J_{0}), \qquad \mathcal{B}_{o}=2b_{o}+4Q(\mathfrak{q}+J_{0}), \ee

\vspace{-0.1cm}
\noi where $J_{0}$ is expressed as
\vspace{-0.1cm}
\begin{align}
J_{0}&=\frac{\mathfrak{q}}{2M^{2}}\bigg(\frac{M^{2}\left[(\sigma^{2}-\Delta_{o})
(R^{2}-4\Delta_{o})-\delta^{2}+4|\mathfrak{q}_{o}|^{2}\right]-4(Qq_{o}+Bb_{o})^{2}}
{4\mathfrak{q}^{2} P_{-}+\delta(\delta-4M\mathfrak{q})}\bigg).
\label{theNUT}\end{align}

\vspace{-0.1cm}
Finally it is not difficult to show that the total angular momentum of the system is expressed in the simple form
\vspace{-0.1cm}
\be 2J=4M\mathfrak{q}-\delta+2J_{0}\left(4M-\frac{\delta}{2\mathfrak{q}}\right).\ee

\vspace{-0.2cm}
Then we have that the sources are two thin rods separated by a coordinate distance $R$ and their location on the symmetry axis depends on the values $\alpha_{1}=-\alpha_{4}=R/2+\sigma$, $\alpha_{2}=-\alpha_{3}=R/2-\sigma$,  as shown in Fig.\ \ref{DK}. As it is well-known, an asymptotically flat spacetime might be considered after killing the NUT charge $J_{0}$, where in this case such a condition is satisfied by means of
\vspace{-0.2cm}
\be \sigma= \sqrt{\Delta- \frac{4\big[|\mathfrak{q}_{o}|^{2}-(Q/M)^{2}q_{o}^{2}\big]-\delta^{2}}{R^{2}-4\Delta}},\qquad \Delta=M^{2}-Q^{2}-\mathfrak{q}^{2}, \label{sigma}\ee

\vspace{-0.1cm}
\noi where we have first eliminated the total magnetic charge from the binary system; i.e., $B=0$. At this point, it is worth noting that each source contains identical magnetic charge with opposite sign. Following the approach of Ref.\ \cite{RMJ}, after the application of straight but non-trivial calculations, eventually one gets the following representation for the Ernst potentials and metric functions \npg

\begin{align}
{\cal{E}}&=\frac{\Lambda+\Gamma}{\Lambda-\Gamma},\quad \Phi=\frac{\chi}{\Lambda-\Gamma},\qquad \Phi_{2}=\frac{F}{\Lambda-\Gamma},\qquad
f=\frac{|\Lambda|^{2}-|\Gamma|^{2}+ |\chi|^{2}}{|\Lambda-\Gamma|^{2}},\qquad \omega=4\mathfrak{q}+\frac{{\rm{Im}}\left[(\Lambda-\Gamma)\overline{\mathcal{G}}-\chi \overline{\mathcal{I}} \right]}{|\Lambda|^{2}-|\Gamma|^{2}+ |\chi|^{2}},\nonu\\
e^{2\gamma}&=\frac{|\Lambda|^{2}-|\Gamma|^{2}+ |\chi|^{2}}{64\sigma^{4}R^{4}\kappa_{o}^{2} r_{1}r_{2}r_{3}r_{4}}, \qquad
\Lambda=2\sigma^{2} \left[R^{2}\kappa_{o}(r_{1}+r_{2})(r_{3}+r_{4})+4a(r_{1}-r_{3})(r_{2}-r_{4})\right]\nonu\\ &+2R^{2}\left[\kappa_{o}(2\Delta-\sigma^{2})-a\right](r_{1}-r_{2})(r_{3}-r_{4})
+2iR\bigg\{\Big(2\mathfrak{q}{\rm Re}(s_{+})+{\rm Im}(p_{+})\Big)\Big[R(\mathfrak{r}_{1}-\mathfrak{r}_{2})(r_{3}-r_{4})\nonu\\
&-2\sigma\big(\mathfrak{r}_{1}r_{4}-\mathfrak{r}_{2}r_{3}+4\sigma r_{3}r_{4}\big)\Big] +\mathfrak{q}\kappa_{o}\Big[ r_{1}\big( R^{2}r_{3}-\kappa_{o}r_{4}\big)-r_{2}\big(\kappa_{o} r_{3}-R^{2}r_{4}\big)-8\sigma^{2}r_{3}r_{4} \Big]\bigg\},\nonu\\
\Gamma&=4\sigma R\left(M\Gamma_{o}- b\chi_{+}\right),\quad F=(4\mathfrak{q}+iz)\chi-i\mathcal{I},\quad \chi=-4\sigma R\left(Q\Gamma_{o}+2\mathbb{Q}\chi_{+}\right),\quad \Gamma_{o}=R \chi_{-}-2\sigma \chi_{s}+2\chi_{1+}, \nonu\\
\mathcal{G}&= 2z\Gamma+  8\sigma^{2}\bigg\{ R\Big[2\big({\rm Re}(a)-2|\mathfrak{q}_{o}|^{2}\big)
+Q^{2}\kappa_{o}\Big](r_{1}r_{2}-r_{3}r_{4})+2i\mathfrak{q}R^{2}\kappa_{o}(r_{2}r_{3}+r_{1}r_{4})+2i\Big[R{\rm Im}(a)+Q\xi_{0}-4\mathfrak{q}|\mathfrak{q}_{o}|^{2}\Big] \nonu\\
&\times (r_{1}-r_{3})(r_{2}-r_{4})\bigg\}-4R^{2}\bigg\{\sigma\left[2a-(R-2\sigma)
\Big(2(R+2i\mathfrak{q})s_{+}+p_{+} \Big)\right]+i\left(Q\xi_{o}+2Qb_{o}\kappa_{o}-4\mathfrak{q}|\mathfrak{q}_{o}|^{2}\right)\bigg\}
(r_{1}-r_{2})(r_{3}-r_{4})\nonu\\
&+2\sigma R\bigg\{4R\Big(2\kappa_{o} \Delta -{\rm Re}(a)\Big)r_{4}+\Big[Q(4q_{o}+QR)\kappa_{o}+4R|\mathfrak{q}_{o}|^{2}\Big](r_{3}+r_{4})\bigg\}(r_{1}-r_{2})\nonu\\
&+2\sigma R \bigg\{4R\Big(2\kappa_{o} \Delta -{\rm Re}(a)\Big)r_{2}-\Big[Q(4q_{o}-QR)\kappa_{o}-4R|\mathfrak{q}_{o}|^{2}\Big](r_{1}+r_{2})\bigg\}(r_{3}-r_{4})
+4M\sigma R\big(\kappa_{o}\chi_{+} +2R\chi_{1-}+4\sigma\chi_{p}\big)\nonu\\
&-4b\sigma R(R\chi_{-}+2\sigma\chi_{s})-8\sigma R(Qb+2M\mathbb{Q})\Big[ 2\bar{\mathfrak{q}}_{o}\big(\mathfrak{r}_{1}-\mathfrak{r}_{2}+\mathfrak{r}_{3}-\mathfrak{r}_{4}\big)+Q\kappa_{o}(r_{1}-r_{2}-r_{3}+r_{4})\Big],\nonu\\
\mathcal{I}&=A\Big[4\sigma^{2}(r_{1}-r_{3})(r_{2}-r_{4})-R^{2}(r_{1}-r_{2})(r_{3}-r_{4})\Big]
+R\kappa_{-}\Big[B_{+}\kappa_{o}r_{1}-B_{-}R\mathfrak{r}_{2} \Big]r_{4}+R\kappa_{+}\Big[B_{-}\kappa_{o}r_{2}-B_{+}R\mathfrak{r}_{1} \Big]r_{3}\nonu\\
&-16\sigma^{2}R \Big\{\Big[M(R+2\sigma)(\kappa_{+}+2QR)-B_{+}\mathfrak{q}_{o}\Big]r_{3}r_{4}-R\kappa_{o}(2M\mathbb{Q}+Qb)\Big\}+
8\mathbb{Q}\sigma R(\chi_{1+}+\sigma\chi_{s})\nonu\\
&+ 2\sigma R\Big[Q\big(2R^{2}-8\Delta+\kappa_{o}\big)+8i\mathfrak{q} \mathbb{Q}\Big]\chi_{+}+ 12\sigma R^{2}\mathbb{Q}\chi_{-} + 8Q\sigma R(R\chi_{1-}+2\sigma\chi_{p}), \quad \kappa_{o}=R^{2}-4\sigma^{2}, \nonu\\
\chi_{\pm}&=s_{+}\mathfrak{r}_{1}-s_{-}\mathfrak{r}_{2} \pm
(\bar{s}_{-}\mathfrak{r}_{3}-\bar{s}_{+}\mathfrak{r}_{4}) , \quad \chi_{1\pm}=p_{+}\mathfrak{r}_{1}+p_{-}\mathfrak{r}_{2}\pm
(\bar{p}_{-}\mathfrak{r}_{3}+\bar{p}_{+}\mathfrak{r}_{4}),\quad
\chi_{s}=s_{+}\mathfrak{r}_{1}+s_{-}\mathfrak{r}_{2}+\bar{s}_{-}\mathfrak{r}_{3}+\bar{s}_{+}\mathfrak{r}_{4}, \nonu\\
\chi_{p}&=p_{+}\mathfrak{r}_{1}-p_{-}\mathfrak{r}_{2} + \bar{p}_{-}\mathfrak{r}_{3}-\bar{p}_{+}\mathfrak{r}_{4},\quad
a=(R+2i\mathfrak{q})p_{+}-s_{+}\big[s_{+}-(R+2i\mathfrak{q})^{2}\big], \quad
b=-2q_{o}(Q/M)+i(\delta-4M\mathfrak{q}),\nonu\\
A&=4M\Big[ \Big( 2 \mathbb{Q}+ Q(R-2\sigma)\Big)s_{+}+2Qp_{+}\Big]
+B_{+}\Big[Q\big(R^{2}-4\Delta\big)-2(R+2i\mathfrak{q})\mathfrak{q}_{o} \Big], \quad \kappa_{\pm}=2\mathfrak{q}_{o}-Q(R\pm 2\sigma), \nonu\\
B_{\pm}&=\Big[ R s_{\pm} \pm p_{\pm} +2Q\big(2\bar{\mathfrak{q}}_{o}+Q(R\pm 2\sigma)\big)\Big]/M, \quad
p_{\pm}=-\sigma(R^{2}-4\Delta)\pm i\big[2M\delta+4b_{o}Q-(R+2i\mathfrak{q}){\rm Im}(s_{\pm})\big], \nonu\\ s_{\pm}&=2\Delta \pm \sigma R+ i \mathfrak{q}(R\pm 2\sigma),\quad
\xi_{o}=4Q\Big[M \delta+2b_{o} Q+\mathfrak{q}(\Delta-\sigma^{2})\Big]-(2b_{o}+\mathfrak{q} Q)(R^{2}-4\Delta), \quad \mathbb{Q}=\mathfrak{q}_{o}+2i \mathfrak{q} Q,\nonu\\ 
\mathfrak{r}_{1,4}&=(R-2\sigma)r_{1,4}, \quad
\mathfrak{r}_{2,3}=(R+2\sigma)r_{2,3}, \quad r_{1,2}=\sqrt{\rho^{2}+\left(z-R/2 \mp \sigma\right)^{2}}, \quad
r_{3,4}=\sqrt{\rho^{2}+\left(z+R/2 \mp \sigma\right)^{2}},
\label{sevenparameters}\end{align}

\vspace{-0.1cm}
\noi where Eq.\ (\ref{sevenparameters}) is depicted by a total of seven parameters $\{M,Q,\mathfrak{q},q_{o},b_{o},\delta,R\}$. Notice that Eq.\ (\ref{sevenparameters}) also shows the Kinnersley potential $\Phi_{2}$ \cite{Kinnersley} in order to get directly the magnetic potential $A_{3}$ through
\vspace{-0.1cm}
\be A_{3}={\rm Re}(\Phi_{2})=-4\mathfrak{q} A_{4}-zA'_{3}+{\rm Im} \bigg(\frac{\mathcal{I}}{\Lambda-\Gamma}\bigg). \ee

\vspace{-0.5cm}
\section{Corotating Kerr-Newman binary BHs}
\vspace{-0.3cm}
The above solution Eq.\ (\ref{sevenparameters}) cannot be considered to describe a pair of BHs unless we have been able to solve the axis condition in the middle region among the sources, namely,
\vspace{-0.1cm}
\be \omega\Big(\rho=0, |z|< {\rm{Re}}(\alpha_{2})\Big)=0,\label{omegamiddle}\ee

\vspace{-0.1cm}
\noi where it ensures that both BHs will be apart by a massless strut. The substitution of Eq.\ (\ref{sigma}) into
Eq.\ (\ref{omegamiddle}) will leads us to a quadratic equation for any of the variables $q_{o}$, $b_{o}$ or $\delta$, namely
\vspace{-0.2cm}
\begin{align} &8 \mathfrak{q} P_{0}b_{o}^{2}+2P_{0}(2Q b_{o}+M \delta)(R^{2}-4\Delta)-\big[2\mathfrak{q}s_{o}-(R+2M)\delta\big]
(R^{2}-4\Delta)^{2}+4\mathfrak{q}\Big\{(P_{0}-2s_{o})\Big[2q_{o}^{2}\big(1-2(Q/M)^{2}\big)-\delta^{2}\Big]\nonu\\
&+4s_{o}q_{o}^{2}\Big\}=0, \nonu\\
P_{0}&=(R+2M)^{2}+4\mathfrak{q}^{2}, \quad s_{o}=M(R+2M)-Q^{2},
\label{middle1}\end{align}

\vspace{-0.1cm}
\noi and since this algebraic equation can be readily solved, it is possible to assume for one instant that we know explicitly its solution, thus having two corotating charged sources separated by a massless strut \cite{BachW}. Later on, we will combine Eq.\ (\ref{middle1}) together with the condition that annihilates to each individual magnetic charge with the aim to define corotating Kerr-Newman binary BHs. Thanks to the fact that the strut is massless, the BHs can be surrounded by a Gauss-type law via Komar integrals \cite{Komar}, where the horizon mass of each BH accomplishes its own Smarr formula \cite{Smarr}. On the other hand, in stationary axisymmetric spacetimes Tomimatsu formulas \cite{Tomi} provide us an easy way to calculate straightforwardly Komar conserved quantities in a two-body system once we know a specific metric. Nevertheless, Cl\'ement and Gal'tsov \cite{Galtsov} recently have shown that Tomimatsu formulas are not correct in the presence of magnetic charges, because the mass of the horizon $M_{H}$ suffers contributions coming from a Dirac string joined to the BHs; i.e., each BH is carrying a magnetic flux. Therefore, the corrected Tomimatsu formulas acquire the form \cite{Tomi,Galtsov}
\vspace{-0.2cm}
\begin{align} M_{H}&= -\frac{1}{8\pi}\int_{H} \omega \Psi_{,z}\, d\varphi dz-M_{A}^{S}, \qquad
Q_{H}+iB_{H}=\frac{1}{4\pi i}\int_{H}\omega \Phi_{,z}\, d\varphi dz, \nonu\\
J_{H}&=-\frac{1}{8\pi}\int_{H}\omega\left[1+ \frac{\omega \Psi_{,z}}{2}
-\tilde{A}_{3}A_{3,z}^{'}\right]d\varphi dz -\frac{\omega^{H}M_{A}^{S}}{2},& \label{TomiGaltsov} \end{align}

\vspace{-0.1cm}
\noi where $\Psi={\rm Im} (\cal E)$ and $\tilde{A}_{3}=A_{3}+\omega A_{4}$. Also, $\omega^{H}$ is the value of the metric function $\omega$ over the horizon, while $M_{A}^{S}$ is an extra term related to the presence of the Dirac string, which is given by
\vspace{-0.2cm}
\be M_{A}^{S}=-\frac{1}{4\pi}\int_{H}(A_{3}^{'}A_{3})_{,z}d\varphi dz.\ee

For such a case, the Smarr formula \cite{Smarr} for each BH still holds and reads \cite{Galtsov}
\vspace{-0.2cm}
\be M_{H}=\frac{\kappa S}{4\pi} +2 \Omega J_{H} +\Phi^{H}_{E} Q_{H}=\sigma +2 \Omega J_{H} +\Phi^{H}_{E} Q_{H}, \label{Smarr}\ee

\begin{figure}[ht]
\centering
\vspace{-0.2cm}
\includegraphics[width=6.0cm,height=5.0cm]{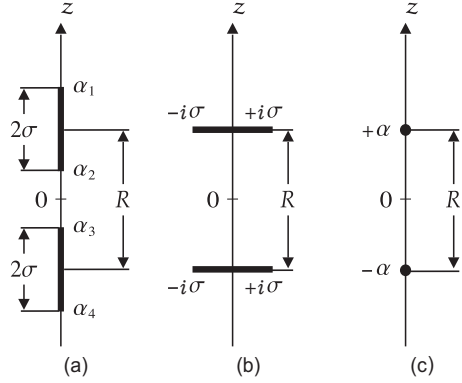}
\caption{Different types of identical Kerr-Newman sources on the symmetry axis: (a) BH configuration $\sigma^{2}>0$; (b) hyperextreme sources if $\sigma \rightarrow i \sigma$ (or $\sigma^{2}<0$ ); (c) the extreme limit case if $\sigma=0$.}
\label{DK}\end{figure}

\vspace{-0.1cm}
\noi where $\kappa$ and $S$ are the surface gravity and the area of the horizon, respectively; both are related to $\sigma$. Furthermore, $\Omega=1/\omega^{H}$ is the angular velocity and $\Phi^{H}_{E}=-A_{4}^{H}-\Omega A_{3}^{H}$ defines the electric potential in the corotating frame of the BH. The aforementioned integrals Eqs.\ (\ref{TomiGaltsov}) are evaluated on the corresponding region that each rod represents the BH horizon [see Fig.\ \ref{DK}(a)]. In this case, one may consider the values of the upper BH horizon: $R/2-\sigma\leq z \leq R/2+\sigma$, $\rho\rightarrow 0$, and $0\leq \varphi \leq 2\pi $, due that the length of both sources is equal. Replacing Eq.\ (\ref{sevenparameters}) into Eq.\ (\ref{TomiGaltsov}), it can be proven that the mass $M_{H}$ and electromagnetic charge $Q_{H}+iB_{H}$ for the upper BH assume the form
\vspace{-0.1cm}
\begin{align} & M_{H}=M+ \frac{2q_{o}(Q/M)P_{0} R(R^{2}-4\Delta)}{\Big[(R+2M)(R^{2}-4\Delta)-4\mathfrak{q}\delta\Big]^{2}+64\mathfrak{q}^{2}q_{o}^{2}(Q/M)^{2}}
-M_{A}^{S},\nonu\\
&Q_{H}+i B_{H}= Q +2\frac{P_{0}(q_{o}+ib_{o})+iQ\Big(\mathfrak{q}(R^{2}-4\Delta)+(R+2M)\big[\delta+2iq_{o}(Q/M)\big]\Big)}
{(R+2M)(R^{2}-4\Delta)-4\mathfrak{q}\big[\delta+2iq_{o}(Q/M)\big]}.
\label{totalcharge}\end{align}

\vspace{-0.1cm}
It is important to note that the individual masses, electric charges, and angular momenta will not be necessarily half of $2M$, $2Q$, and $2J$, respectively, unless the magnetic charges vanish. Moreover, the lower BH will contain a mass $M_{H_(q_{o}\rightarrow -q_{o})}$ and electromagnetic charge $2Q-Q_{H}-iB_{H}$. In the following, we are going to consider two electrically charged models where an absence of magnetic charges is taken into account.

\vspace{-0.5cm}
\subsection{Identically charged Kerr-Newman BHs}
\vspace{-0.3cm}
The first case that is considered here emerges if $q_{o}=0$ and $Q=Q_{H}$, in the formulas of the current section. We find from Eqs.\ (\ref{middle1}) and (\ref{totalcharge}) that an absence of magnetic charges ($B_{H}=0$) is achieved when
\vspace{-0.1cm}
\begin{align}  \delta&=\frac{2\mathfrak{q}(R^{2}-4\Delta)\big[MP_{0}+Q_{H}^{2}(R+2M)\big]}{(R^{2}+2MR+4\mathfrak{q}^{2})P_{0}
+8\mathfrak{q}^{2}Q_{H}^{2}},\qquad
b_{o}=-\frac{\mathfrak{q}Q_{H}(R^{2}-4\Delta)\big(P_{0}+2Q_{H}^{2}\big)}{(R^{2}+2MR+4\mathfrak{q}^{2})P_{0}
+8\mathfrak{q}^{2}Q_{H}^{2}},
\label{sol1}\end{align}

\vspace{-0.1cm}
\noi and therefore $M_{H}=M$ since the term $M_{A}^{S}$ vanishes. For such a situation, the expression for $\sigma$ in the identically charged case is obtainable from Eq.\ (\ref{sigma}); it reads
\vspace{-0.1cm}
\begin{align}  \sigma&=\sqrt{\Delta +\frac{4\mathfrak{q}^{2}(R^{2}-4\Delta)\Big[\big[MP_{0}+Q_{H}^{2}(R+2M)\big]^{2}-Q_{H}^{2}(P_{0}+2Q_{H}^{2})^{2} \Big]}
{\big[(R^{2}+2MR+4\mathfrak{q}^{2})P_{0}+8\mathfrak{q}^{2}Q_{H}^{2}\big]^{2}}},
\label{sigma1}\end{align}

\vspace{-0.3cm}

\vspace{+0.1cm}
\noi whereas the angular momentum of each BH is given by
\vspace{-0.1cm}
\be J_{H}\equiv J=2M\mathfrak{q}-\frac{\delta}{2}= 2M\mathfrak{q}
-\frac{\mathfrak{q}(R^{2}-4\Delta)\big[MP_{0}+Q_{H}^{2}(R+2M)\big]}{(R^{2}+2MR+4\mathfrak{q}^{2})P_{0}
+8\mathfrak{q}^{2}Q_{H}^{2}}. \label{angularmomentum}\ee

\vspace{-0.1cm}
\noi In this binary BH setup, the electric and magnetic dipole moments are specified as $\mathcal{Q}_{o}=0$ and $\mathcal{B}_{o}=2b_{o}+4\mathfrak{q}Q_{H}$. Regarding now the thermodynamical properties contained within the Smarr formula Eq.\ (\ref{Smarr}) we have that $\Omega$ and $\Phi_{E}^{H}$ are
\vspace{-0.2cm}
\begin{align}  \Omega&=\frac{\mathfrak{q}\big[R^{2}-4\Delta+2\sigma(R+2\sigma)\big]-2(2b_{o}Q_{H}+M\delta)}
{\mathcal{L}^{2}+\mathcal{M}^{2}}, \qquad
\Phi_{E}^{H}=Q_{H}\frac{(R+2\sigma)\mathcal{L}-2(b_{o}/Q_{H})\mathcal{M}}
{\mathcal{L}^{2}+\mathcal{M}^{2}},\nonu\\
\mathcal{L}&=MR+2\Delta+(R+2M)\sigma, \quad \mathcal{M}=\delta+\mathfrak{q}(R+2\sigma),
\label{thermoI}\end{align}

\vspace{-0.1cm}
\noi and the area of the horizon $S$ as well as the surface gravity $\kappa$ may be obtained via the formulas \cite{Tomi,Carter}
\vspace{-0.2cm}
\be S= \frac{4\pi \sigma}{\kappa}, \qquad \kappa=\sqrt{-\Omega^{2}e^{-2\gamma^{H}}},\ee

\vspace{-0.1cm}
\noi where $\gamma^{H}$ defines the value that metric function $\gamma$ takes at the horizon. A straightforward calculation yields us to
\vspace{-0.1cm}
\be \frac{S}{4\pi}= \frac{\sigma}{\kappa}=\frac{\mathcal{L}^{2}+\mathcal{M}^{2}}{R(R+2\sigma)}.\ee

\vspace{-0.1cm}
On the other hand, when analyzing the energy-momentum tensor, the line source of pressure and a negative energy density numerically equal to it defines a massless strut \cite{Israel}. Therefore, in order to interpret the pressure exerted on each source, it is necessary to compute the interaction force associated to the strut, which can be obtained in a simple manner with the formula $\mathcal{F}=(e^{-\gamma_{s}}-1)/4$ \cite{Israel,Weinstein}, where $\gamma_{s}$ denotes the constant value that metric function $\gamma$ acquires on the middle region among the sources, thus getting
\vspace{-0.1cm}
\be \mathcal{F}= \frac{\big[(M^{2}-Q_{H}^{2})P_{0}^{2}-4\mathfrak{q}^{2}Q_{H}^{4}\big](P_{0}-8\mathfrak{q}^{2})
-16\mathfrak{q}^{2}Q_{H}^{2}
\big[s_{o}P_{0}-Q_{H}^{4}\big]}{(R^{2}-4\Delta)P_{0}^{3}}.\ee

\vspace{-0.1cm}
Contrary to the vacuum scenario \cite{CCLP}, Eq.\ (\ref{angularmomentum}) cannot be solved analytically because it is leading us to a quintic algebraic equation in the variable $\mathfrak{q}$. So, whenever the numerical analysis should be performed, it is necessary to bear in mind, the physical limits of the solution. In this sense, the minimal distance at which the BH horizons are touching each other (the merger limit) is given by $R_{min}=2\sqrt{M^{2}-Q_{H}^{2}-\mathfrak{q}^{2}}$, while the force $\mathcal{F}\rightarrow \infty$. After taking into account this distance value, we notice from Eqs.\ (\ref{sigma1}) and (\ref{angularmomentum}) that $\sigma=\sqrt{M^{2}-Q_{H}^{2}-(J/2M)^{2}}$ and $\mathfrak{q}=J/2M$, respectively, having the following result
\vspace{-0.2cm}
\be R_{min}=2 \sqrt{M^{2}-Q_{H}^{2}-\left(\frac{J}{2M}\right)^{2}} \equiv 2\sigma, \label{minimaldistance} \ee

\vspace{-0.1cm}
\noi which is leading us to very simple expressions for $\Omega$, $\Phi_{E}^{H}$, $\kappa$, and $S$:
\vspace{-0.1cm}
\begin{align}
\Omega&=\frac{J/M}{4d_{0}},\qquad \Phi_{E}^{H}=\frac{Q_{H}(M+\sigma)}{d_{0}}, \qquad \frac{S}{4\pi}=\frac{\sigma}{\kappa}=2d_{0},\nonu\\
d_{0}&=(M+\sigma)^{2}+(J/2M)^{2},
\label{HorizonpropertiesI}\end{align}

\vspace{-0.1cm}
\noi where we have employed the fact that the parameters $b_{o}$ and $\delta$ are equal to zero during the merger limit. Another physical limit to be considered is when $R \rightarrow \infty$, representing the physical case in which the sources move far away from each other and the interaction force $\mathcal{F}\rightarrow 0$. It is quite easy to show from Eqs.\ (\ref{sol1})-(\ref{angularmomentum}) that $\mathfrak{q}=J/M$ and $b_{o}=-Q_{H} J/M$, and thereby one gets the thermodynamical features for a single Kerr-Newman BH, namely \npg
\vspace{-0.2cm}
\begin{align}
\Omega&=\frac{J/M}{d_{1}},\qquad \Phi_{E}^{H}=\frac{Q_{H}(M+\sigma)}{d_{1}}, \qquad \frac{S}{4\pi}=\frac{\sigma}{\kappa}=d_{1},\nonu\\
d_{1}&=(M+\sigma)^{2}+(J/M)^{2}, \quad \sigma=\sqrt{M^{2}-Q_{H}^{2}-(J/M)^{2}},
\label{HorizonpropertiesII}\end{align}

\vspace{-0.1cm}
\noi where one confirms that the size of the BH horizon $2\sigma$ during the merger limit is bigger in comparison to the isolated case, but its thermodynamical properties decrease their corresponding values. This statement is in agreement with the Smarr formula. At large distances the interaction force acquires the approximate value
\vspace{-0.1cm}
\begin{align} \mathcal{F}&\simeq\frac{M^{2}-Q_{H}^{2}}{R^{2}}\Bigg[1+\frac{4\big[M^{2}
-Q_{H}^{2}-3(J/M)^{2}\big]}{R^{2}} +\frac{8 (J/M)^{2}\big[10M^{4}-15M^{2}Q_{H}^{2}+3Q_{H}^{4}\big]}{M(M^{2}-Q_{H}^{2})R^{3}}+O \left(\frac{1}{R^{4}} \right)\Bigg].\label{asymptoticforce} \end{align}

\vspace{-0.5cm}
\subsection{Oppositely charged Kerr-Newman BHs}
\vspace{-0.3cm}
The second electrically charged model comes to light immediately when the total electric charge
is eliminated from the binary system by doing now $Q=0$. Then, the magnetic charges can be removed from the solution only if $b_{o}=0$ and
\vspace{-0.2cm}
\begin{align}  \delta&=\frac{2\mathfrak{q}(R^{2}-4M^{2}+4\mathfrak{q}^{2})\big[MP_{0}-Q_{H}^{2}(R+2M)\big]}
{(R^{2}+2MR+4\mathfrak{q}^{2})P_{0}
-8\mathfrak{q}^{2}Q_{H}^{2}},\qquad
q_{o}=\frac{Q_{H}R(R^{2}-4M^{2}+4\mathfrak{q}^{2})P_{0}}{2\Big[(R^{2}+2MR+4\mathfrak{q}^{2})P_{0}
-8\mathfrak{q}^{2}Q_{H}^{2}\Big]},
\label{sol2}\end{align}

\vspace{-0.2cm}
\noi where Eqs.\ (\ref{middle1}) and (\ref{totalcharge}) are identically fulfilled with these expressions. Notice that once again $M_{H}=M$ and $M_{A}^{S}=0$, where now the electric and magnetic dipole moments are $\mathcal{Q}_{o}=2q_{o}$ and $\mathcal{B}_{o}=0$. In the oppositely charged case $\sigma$ takes the form
\vspace{-0.2cm}
\begin{align}  \sigma&=\sqrt{M^{2}-\mathfrak{q}^{2} +\frac{(R^{2}-4M^{2}+4\mathfrak{q}^{2})\Big[
\big[2\mathfrak{q}\big(MP_{0}-Q_{H}^{2}(R+2M)\big)\big]^{2}-(Q_{H}RP_{0})^{2}\Big]}
{\big[(R^{2}+2MR+4\mathfrak{q}^{2})P_{0}-8\mathfrak{q}^{2}Q_{H}^{2}\big]^{2}}},
\label{sigma2}\end{align}

\vspace{-0.2cm}
\noi and the angular momentum is expressed by means of another quintic algebraic equation
\vspace{-0.1cm}
\be J=2M\mathfrak{q}-\frac{\mathfrak{q}(R^{2}-4M^{2}+4\mathfrak{q}^{2})\big[MP_{0}-Q_{H}^{2}(R+2M)\big]}
{(R^{2}+2MR+4\mathfrak{q}^{2})P_{0}
-8\mathfrak{q}^{2}Q_{H}^{2}}. \label{angularmomentumopp}\ee

\vspace{-0.1cm}
It follows that we have obtained a \emph{corotating black dihole model} which in the lack of rotation ($\mathfrak{q}=0$) specifies a non-extreme Emparan's dihole \cite{Emparan}. With regard to the thermodynamical characteristics we have now that $\Omega$, $\Phi_{E}^{H}$, $\kappa$, and $S$ are simplified as
\vspace{-0.2cm}
\begin{align}  \Omega&=\frac{\mathfrak{q}\big[R^{2}-4M^{2}+4\mathfrak{q}^{2}+2\sigma(R+2\sigma)\big]-2M\delta}
{\mathcal{N}^{2}+\mathcal{M}^{2}}, \qquad
\Phi_{E}^{H}=Q_{H}\frac{2(q_{o}/Q_{H})\mathcal{N}}
{\mathcal{N}^{2}+\mathcal{M}^{2}}, \qquad \frac{S}{4\pi}= \frac{\sigma}{\kappa}=\frac{\mathcal{N}^{2}+\mathcal{M}^{2}}{R(R+2\sigma)}, \nonu\\
\mathcal{N}&=MR+2M^{2}-2\mathfrak{q}^{2}+(R+2M)\sigma,
\label{thermoII}\end{align}

\vspace{-0.1cm}
\noi while the formula of the force reads
\vspace{-0.2cm}
\begin{align} \mathcal{F}&= \frac{\big(M^{2}P_{0}^{2}-4\mathfrak{q}^{2}Q_{H}^{4}\big)(P_{0}-8\mathfrak{q}^{2})+Q_{H}^{2} P_{0}\Big[R^{2}P_{0}-4\mathfrak{q}^{2}(R^{2}-4M^{2}+4\mathfrak{q}^{2})\Big]}{(R^{2}-4M^{2}+4\mathfrak{q}^{2})P_{0}^{3}}.
\label{force2}\end{align}

\vspace{-0.1cm}
Similar to the situation with identical electric charges, it follows that the merger limit now results to be
\vspace{-0.1cm}
\be R_{min}=2 \sqrt{M^{2}-\left(\frac{J}{2M}\right)^{2}} \equiv 2\sigma, \label{TheRII}\ee

\vspace{-0.1cm}
\noi where $\mathfrak{q}=J/2M$. In this limit value of the distance the properties on the horizon are written down as follows
\vspace{-0.2cm}
\begin{align}
\Omega&=\frac{J/M}{4d_{0}},\qquad \Phi_{E}^{H}=0, \qquad \frac{S}{4\pi}=\frac{\sigma}{\kappa}=2d_{0},
\label{HorizonpropertiesIII}\end{align}

\vspace{-0.1cm}
\noi from which it is shown that the electric potential vanishes, and therefore, the event horizon $2\sigma$ contains the same length as in the vacuum case \cite{Costa,CCLP}. Furthermore, in the other limit $R \rightarrow \infty$, we have that $\mathfrak{q}=J/M$ and the electric dipole behaves as $\mathcal{Q}_{o} \sim Q_{H}R$, where it is possible to recover the description of one isolated Kerr-Newman BH when deriving exactly the same formulas described above in Eq.\ (\ref{HorizonpropertiesII}). Finally, the force at infinite separation distance contains the next behavior
\vspace{-0.2cm}
\begin{align} \mathcal{F}&\simeq\frac{M^{2}+Q_{H}^{2}}{R^{2}}\Bigg[1-\frac{4MQ_{H}^{2}}{M^{2}+Q_{H}^{2}}
\Bigg(\frac{1}{R}-\frac{3M}{R^{2}} \Bigg)+\frac{4\big[M^{2}
-3(J/M)^{2}\big]}{R^{2}} 
+\frac{8 \Big[(J/M)^{2}\big(10M^{4}+19M^{2}Q_{H}^{2}+3Q_{H}^{4}\big)-6M^{4}Q_{H}^{2} \Big]}{M(M^{2}+Q_{H}^{2})R^{3}}\nonu\\
&+O \left(\frac{1}{R^{4}} \right)\Bigg].\label{asymptoticforce} \end{align}

\begin{figure}[ht]
\centering
\includegraphics[width=8.5cm,height=5.0cm]{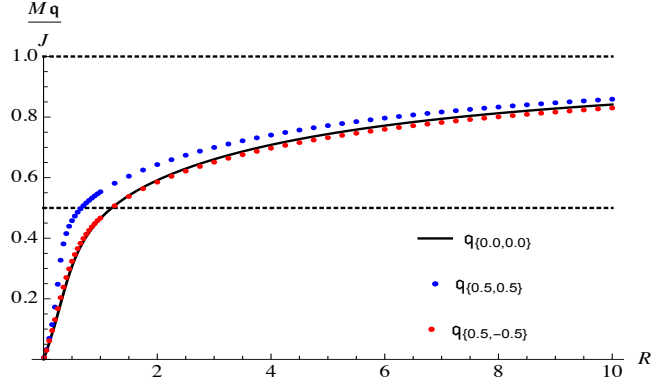}
\vspace{-0.2cm}
\caption{Behavior of the parameter $\mathfrak{q}$ in the non-extreme situation, taking the values $M=1$, $J_{H}=1.6$, and $Q_{H}=0.5$. The identically/oppositely charged case is denoted with the same/contrary signs inside the brackets. Also, the vacuum scenario is indicated by $Q_{H}=0.0$.}
\label{Theq}\end{figure}

\vspace{-0.1cm}
The physical values for the variable $\mathfrak{q}$ earlier discussed for the non-extreme case are shown in Fig.\ \ref{Theq}. 

\vspace{-0.3cm}

\vspace{-0.5cm}
\subsection{The extreme case of BHs}
\vspace{-0.3cm}
The extreme limit of Eq.\ (\ref{sevenparameters}) can be achieved once is established $\sigma=0$, where such a metric may be expressed in a quite simple form by using the Perjes factor structure \cite{Perjes}, thus having
\vspace{-0.1cm}
\begin{align} {\cal{E}}&=\frac{\Lambda-\Gamma}{\Lambda+\Gamma},\quad \Phi=\frac{\chi}{\Lambda+\Gamma}, \quad \Phi_{2}=\frac{(4\mathfrak{q}+i\alpha x y)\chi-i \mathcal{I}}{\Lambda+\Gamma},\quad
f=\frac{\mathcal{D}}{\mathcal{N}}, \quad \omega=\frac{\alpha(x-1)(y^{2}-1)\mathcal{W}}{\mathcal{D}},\quad e^{2\gamma}=\frac{\mathcal{D}}{\alpha^{8}(x^{2}-y^{2})^{4}}, \nonu\\
\Lambda&=\alpha^{2}\left[p(x^{2}-y^{2})^{2}+\Delta(x^{4}-1)\right] +(\mathfrak{q} r-\Delta p)(y^{4}-1)+2i\alpha xy\Big[(r+\mathfrak{q}\alpha^{2})(y^{2}-1)-\alpha^{2} \mathfrak{q}(x^{2}+y^{2}-2)\Big], \nonu\\
\Gamma&=2M\mathbb{P}_{1}-\epsilon \mathbb{P}_{2}, \quad
\chi= 2(Q \mathbb{P}_{1}+\mathfrak{q}_{o}\mathbb{P}_{2}),\nonu\\
\mathcal{I}&=-\alpha^{2}\bigg\{2\mathfrak{q}_{o}\Big[M(1-y^{2})+4i\mathfrak{q}y\Big] +Q\Big[2\alpha M_{o}x+\bar{\epsilon}(1+y^{2})
+2\big[\alpha(2Mx+\alpha)-2\mathfrak{b}_{o}y-2p\big]y\Big] \bigg\}(x^{2}-1)\nonu\\
&-\bigg\{\mathfrak{q}_{o}\Big[(2Mp-i\mathfrak{q}\bar{\epsilon})(1+y^{2})-2\alpha x\big[4(\alpha M x+\Delta)-\bar{\epsilon} y\big]+4i(\alpha^{2} \mathfrak{q}-r)y\Big]\nonu\\
&+Q\Big[(p\bar{\epsilon}-2iMr)(1+y^{2})-2\alpha x\big[\alpha \epsilon x-\bar{M}_{o}p+i(2r-\alpha^{2} \mathfrak{q})\big]-2\big[(\alpha^{2}+2\Delta)p+iM_{o}r\big]y\Big]\bigg\}(1-y^{2})
+2(\mathfrak{q}_{o}+Q\bar{M}_{o})\mathbb{P}_{1},\nonu\\
\mathbb{P}_{1}&=\alpha^{3} x(x^{2}-1)+(\alpha p x-i r y)(1-y^{2}),\qquad
\mathbb{P}_{2}= \alpha^{2}y (x^{2}-1)+(py-i\mathfrak{q}\alpha x)(1-y^{2}),\nonu\\
\mathcal{N}&= \mathcal{D}+ \Theta \Pi-(1-y^{2})(x-1)\Sigma {\rm T},
\qquad \mathcal{D}= \Theta^{2}+(x^{2}-1)(y^{2}-1)\Sigma^{2}, \qquad
\mathcal{W}=(x+1)\Sigma\Pi-\Theta {\rm T},  \nonu\\
\Theta&=\alpha^{2}\Big[p(x^{2}-y^{2})^{2}+\Delta(x^{2}-1)^{2}\Big] +(\mathfrak{q} r-\Delta p)(y^{2}-1)^{2}, \qquad \Sigma=2\alpha\Big(\alpha^{2} \mathfrak{q} x^{2}-ry^{2} \Big), \nonu\\
\Pi&= 2\alpha x\Big[2\alpha^{2}M(x^{2}-y^{2})+2\alpha(2M^{2}-Q^{2})x 
+(2M\Delta+\mathfrak{q} \delta)(1+y^{2})\Big]
-2y\Big\{ \mathfrak{b}_{o}\big[\alpha^{2}(x^{2}-y^{2})+2\alpha M x \nonu\\
&+\Delta(1+y^{2})-\mathfrak{b}_{o}y\big]-(\delta^{2}-2|\mathfrak{q}_{o}|^{2})y\Big\},\qquad
{\rm T}=2\Big[\alpha\big[ \mathfrak{a}_{o}+\delta(\alpha x+M)+\mathfrak{q}\mathfrak{b}_{o}y\big](1+x)+(2Mr+p\delta)(1-y^{2}) \Big],\nonu\\
r&=\mathfrak{a}_{o}-\mathfrak{q}(p-\Delta), \quad p=\alpha^{2}-\Delta, \quad \epsilon=\mathfrak{b}_{o}+i\delta,\quad
\mathfrak{a}_{o}=M\delta+2b_{o}Q, \quad \mathfrak{b}_{o}=-2q_{o}(Q/M), \quad M_{o}=M+i\mathfrak{q},
\label{extremeflat}\end{align}

\vspace{-0.1cm}
\noi where $(x,y)$ are prolate spheroidal coordinates denoted as
\vspace{-0.2cm}
\be x=\frac{r_{+}+r_{-}}{2\alpha}, \qquad y=\frac{r_{+}-r_{-}}{2\alpha}, \qquad r_{\pm}=\sqrt{\rho^{2} + (z \pm \alpha)^{2}}.   \label{prolates}\ee

\begin{figure}[ht]
\centering
\includegraphics[width=8.5cm,height=5.0cm]{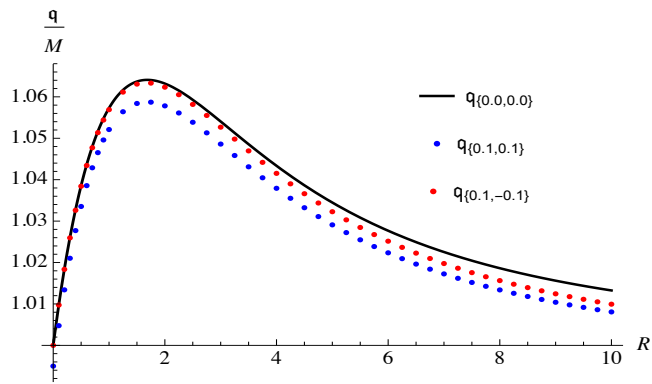}
\vspace{-0.2cm}
\caption{The parameter $\mathfrak{q}$ in the extreme case for fixed mass $M=1$ and electric charge $Q_{H}=0.1$.  }
\label{Theqext}\end{figure}

\begin{figure}[ht]
\centering
\includegraphics[width=8.5cm,height=5.0cm]{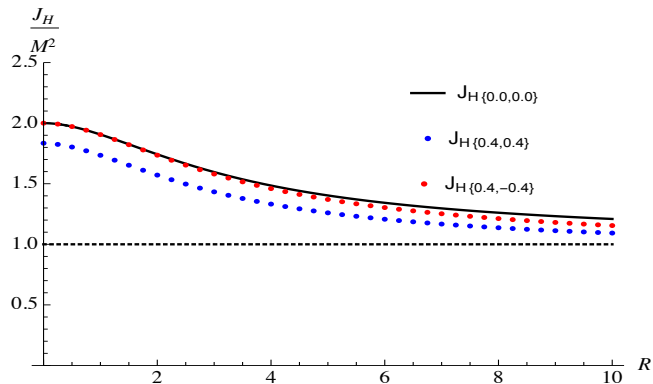}
\vspace{-0.2cm}
\caption{The angular momentum in the extreme case for $M=1$ and electric charge $Q_{H}=0.4$.}
\label{Jext}\end{figure}

\vspace{-0.1cm}
Therefore, this metric allows to treat identically charged extreme Kerr-Newman BHs once is established first $q_{o}=0$, and $Q=Q_{H}$ on it. The explicit values of the angular momentum will depend on the parameter $\mathfrak{q}$ after solving Eq.\ (\ref{sigma1}) for the condition $\sigma=0$. Typical shapes are shown in Figs.\ \ref{Theqext} and \ref{Jext}. Let us now consider the extreme case of BHs during the merger limit, with the goal to derive non-trivial expressions for the force and area of the horizon at this particular distance value. We notice from Eq.\ (\ref{minimaldistance}) that the minimal distance is given by $R_{min}=0$ from which it is possible to get
\vspace{-0.2cm}
\be J=2M \sqrt{M^{2}-Q_{H}^{2}},\ee

\vspace{-0.1cm}
\noi where $\mathfrak{q}=\sqrt{M^{2}-Q_{H}^{2}}$. For such a situation, the merger process will start to form a single extreme Kerr-Newman BH of total mass $M_{T}=2M$, total electric charge $Q_{T}=2Q_{H}$, and total angular momentum $J_{T}=2J$, fulfilling a well-known relation, namely
\vspace{-0.2cm}
\be J_{T}=M_{T}\sqrt{M_{T}^{2}-Q_{T}^{2}}.\ee 

\vspace{-0.1cm}
So, in order to compute the final values that the force and horizon area acquire in the physical scenario where both extreme BHs touch to each other (but they do not merge yet!), we are going to establish initially $\sigma=0$ in Eq.\ (\ref{sigma1}), and later on, we will apply Taylor's expansion around $R=0$ by using $\mathfrak{q}=\sqrt{M^{2}-Q_{H}^{2}}+ C_{0}R$, with the aim to calculate a first order contribution in the whole set of physical and thermodynamical properties. With this procedure, the result turns out to be
\vspace{-0.2cm}
\begin{align} C_{0}&=\frac{-4+3x^{2}+ \delta_{1}}{4(4-x^{2})\sqrt{1-x^{2}}}, \quad x:=\frac{|Q_{H}|}{M}<1,\nonu\\
\delta_{1}&=\varepsilon \sqrt{(2-x^{2})(16-16x^{2}+x^{4})}, \quad \varepsilon=\pm 1, \end{align}

\vspace{-0.1cm}
\noi whereas the force and area of the horizon are given explicitly by
\vspace{-0.2cm}
\begin{align} \mathcal{F}&=\frac{16-24x^{2}+3x^{4}+3x^{6}+2(4-3x^{2})
\delta_{1}}{16(2-x^{2})^{3}},\qquad
S=4\pi M^{2}(2-x^{2})\bigg[1+ \bigg(\frac{4-3x^{2}-\delta_{1}}{(4-x^{2})\sqrt{1-x^{2}}}\bigg)^{2} \bigg].\end{align}

\vspace{-0.1cm}
The angular velocity and electric potential are obtained from Eq.\ (\ref{HorizonpropertiesI}) by doing simply $\sigma=0$ and these properties do not require to much special attention from us. It should be pointed out that there exist two states during the merger limit depending on the sign of $\varepsilon$. In the first/second case; i.e., when $\varepsilon=+1/-1$, the force is positive/negative and therefore attractive/repulsive, where the area of the horizon is smaller in the attractive case compared to the repulsive case. In addition, in the absence of electric charge $Q_{H}=0$ one gets the following formulas
\vspace{-0.2cm}
\be \mathcal{F}=\frac{1+2\varepsilon \sqrt{2}}{8}, \qquad S=16\pi(2-\varepsilon \sqrt{2})M^{2}, \label{forceareavacuum}\ee

\vspace{-0.1cm}
\noi where it can be seen that only the attractive case has been considered earlier in \cite{Ciafre-Rodriguez}.\footnote{By applying the same limiting procedure in the unequal vacuum case \cite{Cabrera2018, Cabrera2018PTP}, one might be able to obtain the non-identical version of Eq.\ (\ref{forceareavacuum}), namely
\vspace{-0.1cm}
\begin{align} \mathcal{F}&=\frac{M_{1}M_{2}+ \varepsilon \sqrt{ 2M_{1}M_{2}}(M_{1}+M_{2})}{2(M_{1}+M_{2})^{2}}, \qquad S_{i}=8\pi M_{i}(M_{1}+M_{2})^{2}\bigg(\frac{M_{1}+M_{2}-\varepsilon \sqrt{2 M_{1}M_{2}}}{M_{1}^{2}+M_{2}^{2}}\bigg), \quad i=1,2, \end{align}
\noi where $M_{1}=M_{2}=M$, recovers it.} 

When is now placed $Q=0$ and $b_{o}=0$ in Eq.\ (\ref{extremeflat}) one may treat oppositely charged extreme Kerr-Newman BHs. In the same manner as in the identically charged situation, the angular momentum adopts the values from the parameter $\mathfrak{q}$ that fulfill the condition $\sigma=0$ [see Figs.\ \ref{Theqext} and \ref{Jext}]. Moreover, in the merger limit of extreme oppositely charged BHs, the same limiting procedure described before might be also included here. In this respect, we have from Eq.\ (\ref{TheRII}) that $R_{min}=0$ is valid only if the angular momentum and mass are related by means of $J=2M^{2}$, while now $\mathfrak{q}=M$. After carrying out once again Taylor's expansion around $R=0$, eventually the interaction force and area of the horizon are expressed as
\vspace{-0.2cm}
\begin{align} \mathcal{F}&=\frac{16+8x^{2}+3x^{4}+2 \sqrt{2}\varepsilon (4+x^{2})\sqrt{16+x^{4}}}{128},\qquad
S=8\pi M^{2}\bigg[1+ \bigg(\frac{4+x^{2}-\varepsilon \sqrt{2} \sqrt{16+x^{4}}}{4-x^{2}}\bigg)^{2} \bigg],\end{align}

\vspace{-0.1cm}
\noi where according to the sign taken by $\varepsilon$ there exist two states that also behave exactly in the same manner as in the identically charged case. These expressions are reduced to the aforementioned Eq.\ (\ref{forceareavacuum}) if the electric charges are not present.

\vspace{-0.3cm}
\vspace{-0.3cm}

\vspace{-0.5cm}
\section{Conclusion}
\vspace{-0.4cm}
In this paper, we have worked out a physical metric that permits us the study of identical corotating Kerr-Newman binary BH models in which the sources contain equal (or opposite) electric charges. These clearly extend the earlier results provided in Refs. \cite{Costa,CCLP}. In order to derive these $4$-parametric physical models, the axis condition in between sources has been combined with the one eliminating the magnetic charges; therefore, the sources are two Kerr-Newman BHs supported by a conical singularity. Our suitable parametrization allowed us to get concise formulas for the physical and thermodynamical features of the BHs, after getting first the half-length parameter $\sigma$ which represents the BH horizon in cylindrical coordinates. On one hand, in both models the horizon mass $M_{H}=M$ due to the fact that magnetic charges are not involved. On the other hand, it is quite clear that $M_{H} \neq M$ if there exists a Dirac string joined to the BHs \cite{Galtsov,Clement}. The contribution of the string mass $M_{A}^{S}$ clearly deserves further research. We believe that the whole thermodynamical components that are being part of the Smarr formula in each corotating charged model are useful to complement the recent results of \cite{CMRV}, where the notion of thermodynamic length \cite{AGK,KZ} has been considered to study the first law of thermodynamics in binary systems of equal counterrotating Kerr-Newman BHs.

In the extreme limit of BHs we have introduced a binary metric with a quite simple aspect. Remarkably, our physical treatment led us to derive in both models two final states during the merger limit, in which the force is attractive/repulsive while the horizon area is smaller in the attractive case in contrast to the repulsive scenario. It is presumable that such an extreme metric will be helpful to extend the previous results included in Ref. \cite{Ciafre-Rodriguez} on a near extreme binary BH geometry, with a more physical aspect. To conclude, we would like to mention that the most satisfactory exact solution as describing unequal configurations is extremely complicated, but we do not exclude that after some efforts this problem might be solved by following the approach considered within this work.

\vspace{-0.5cm}\section*{Acknowledgements}
\vspace{-0.4cm} We thank the referee for his valuable remarks and suggestions. ICM acknowledges the financial support of SNI-CONACyT, M\'exico, grant with CVU No. 173252.

\end{document}